\documentstyle[pra,aps,epsf,twocolumn]{revtex}

\begin{document}
\draft
\title{Volume of the set of mixed entangled states II}
\author{Karol \.Zyczkowski}
\address{Instytut Fizyki im. Mariana Smoluchowskiego,\\
Uniwersytet Jagiello{\'n}ski,\\
ul. Reymonta 4, 30-059 Krak{\'o}w, Poland}
\date{\today}
\maketitle

\begin{abstract}
The problem of how many entangled or, respectively, separable states
there are in the set of all quantum states is investigated. We study to what
extent the choice of a measure in the space of density matrices
$\varrho$
describing $N$--dimensional quantum systems affects the results
obtained. We
demonstrate that the link between the purity of the mixed states and the
probability of entanglement is not sensitive to the measure chosen. Since
the criterion of partial transposition is not sufficient to distinguish all
separable states for $N\ge 8$, we develop an efficient algorithm to
calculate numerically the entanglement of formation of a given mixed quantum
state, which allows us to compute the volume of separable states for $N=8$
and to estimate the volume of the bound entangled states in this case.
\end{abstract}


%\narrowtext

\section{Introduction}

{\sl Entangled states} have been known almost from the very beginning of
quantum
mechanics and their somewhat unusual features have been investigated for
many years. However, the recent developments of the theory of quantum
information and quantum computing caused a rapid increase of the interest in
studying their properties and possible applications. To illustrate this
trend let us quote some data from the Los Alamos quantum physics archives.
In 1994 only one paper posted in these archives contained the key word
'entangled' (or entanglement) in the title, while two such papers were
posted in 1995. Since then the number of such papers
has increased dramatically,
and was equal to $8$, $30$ and $70$, in the consecutive years 1996, 1997 and
1998, respectively.

We do not dare to fit some fast growing curves to these data nor to
speculate, when such an increase will eventually saturate. On the other
hand, since so many authors have dealt with  entangled states, it is
legitimate to ask, whether such states are 'typical' in quantum theory, or
if they are rather rare and unusual. Vaguely speaking, we shall be
interested in the relative likelihood of encountering an entangled state
\cite{ZHSL}. One may also ask a complementary question concerning the
set of the {\sl separable} states, which can be represented as a sum of
product states.

Consider a quantum system described by the density matrix $\rho$
which represents a mixture of the pure states of the $N$
dimensional Hilbert space. Let us assume that the system consists of two
subsystems, of dimension $n_A$ and $n_B$, where $N=n_A n_B$. To formulate
the basic question {\sl 'What is the probability of finding an entangled
state of size $N$?'}, one needs to:

(i) define the probability measure $\mu$, according to which the random
density matrices $\rho$ are drawn,

(ii) find an efficient technique, which would allow one to judge, whether  a
given mixed state $\rho$ is entangled.

Representing any density matrix in
 the diagonal form $\rho =U dU^{\dagger}$
we proposed \cite{ZHSL} to use a product measure $\mu =\Delta_1 \times
\nu$,
where $\Delta_1$ describes the uniform measure on the simplex $%
\sum_{i=1}^{N}d_{i}=1$ and $\nu $ stands for the Haar measure in the space
of unitary matrices $U(N)$. Based on the partial transposition criterion
\cite{Pe96,Ho1} we found that under this measure the volume of the set of
separable states is positive and decreases with the system size $N$ \cite
{ZHSL}. Some more general analytical bounds were also provided by Vidal and
Tarrach \cite{VT98}. Recently Slater suggested estimating the same quantity
using some other measures in the space
of the density matrices \cite{Sl1,Sl2}.
One may thus expect that the volume of the separable states
depends on the measure chosen. We show that this is indeed the case.
In this work
 we investigate which statistical properties describing the set of
the entangled states may be universal, e.g. do not depend on the measure
used.  In particular  we
demonstrate that the relation between the purity of mixed states and the
probability of entanglement is not very sensitive to the measure assumed.
 Basing on numerical results  we conjecture  that the volume of the
separable states decreases exponentially with the system size $N$.

For $N=4$ and $N=6$ a density matrix is separable if and only if its partial
transpose is positive \cite{Ho1}. For $N>6$, however, there exist states,
which are not separable and do satisfy this criterion \cite{Ho2}. These
states cannot be distilled to the singlet form and are
 called {\sl bound entangled} \cite{Ho3,Ho98,VMSST,LP98}.
 Since there are no explicit conditions
allowing one to distinguish between separable and bound entangled states, in
Ref. \cite{ZHSL} only the upper bound for the volume of separable states was
considered for $N>6$. In this paper we present an efficient numerical
method of
computing the {\sl entanglement of formation} $E$ \cite{BVSW96} for any
density matrix. This method allows us to estimate the volume of bound
entangled states, by taking a reasonably small cut-off entanglement $E_c$
and counting these states satisfying the partial transposition criterion for
which $E>E_c$. Our numerical results are to large extent independent on
the exact value of $E_c$.

The paper is organized as follows. In section II we review the necessary
definitions and study how the upper bound of the volume of the separable
states depends on the system size and the measure used. The next section is
devoted to analysis of the simplest case $N=4$, for which the bound
entangled states do not exist. In this case the analytical formula for the
entanglement of formation is known \cite{HW97,Wo98} and we study how
this quantity
changes with the purity of the mixed states.  In section IV we study the
case $N=8$ and estimate the volume of the free entangled states, bound
entangled states and separable states. The paper is concluded by Section V
containing a list of open questions. In Appendix A we prove the rotational
invariance of the two distinguished measures $\Delta_o$ and $\Delta_u$,
defined on the $N-1$ dimensional simplex and demonstrate the link with the
ensembles of random matrices. The algorithm of computing the entanglement of
formation for a given density matrix is presented in Appendix B.

\section{Volume of the states with positive partial transposition}

\subsection{Product measures in the space of mixed density matrices}

To discuss the probability, that a mixed quantum state possesses a given
property, one needs to define a probability measure $\mu $ in the space of
density matrices ($N\times N$ positive Hermitian matrices with trace equal
to unity).  Each density matrix can be diagonalized by a unitary
rotation.  Let $B$ be a diagonal unitary matrix. Since
\begin{equation}
\rho =UdU^{\dagger }=UBdB^{\dagger }U^{\dagger },  \label{diag2}
\end{equation}
therefore the rotation matrix $U$ is determined up to $N$ arbitrary
phases entering $B$. The total number of independent variables
used to parametrize in this way any density matrix $\rho$ is equal to
$N^2-1$.
Since the literature seemed not to distinguish any natural
measure in this space, we approached the problem by defining a product
measure \cite{ZHSL}
\begin{equation}
\mu _{u}=\Delta _{1}\times \nu _{H}.  \label{muu}
\end{equation}
The measure $\nu $ is defined in the space of unitary matrices $U(N)$, while
$\Delta $ is defined in the $(N-1)$-dimensional simplex determined by the
trace condition $\sum_{i=1}^{N}d_{i}=1$. In Ref. \cite{ZHSL} we took for $%
\nu $ the Haar measure on $U(N)$, while the uniform measure $\Delta_1$
was used on the simplex. Our choice was motivated by the fact that the
both component measures are rotationally invariant. For $\nu _{H}$ this
follows directly from the definition of the Haar measure, while in Appendix
A we prove that the uniform measure $\Delta _{1}$ corresponds to taking for
the vector $d_{i}$ the squared moduli of complex elements of a column or a
row (say, the first column) of an auxiliary random unitary matrix $V$ drawn
with respect to $\nu _{H}$
\begin{equation}
d_{i}=|V_{i1}|^{2}.  \label{dii}
\end{equation}
In the sequel we will thus refer to the measure defined by (\ref{muu}) as
the unitary product measure $\mu _{u}$.

As correctly pointed out by Slater \cite{Sl1,Sl2}, our choice of the measure
is by far not the only one possible. He discussed several possible measures,
and proposed to pick the measure on the $(N-1)$D simplex from a certain
family of the Dirichlet distributions
\begin{equation}
\Delta_{\lambda}(d_1,\dots,d_{N-1}) = C_{\lambda} d_1^{\lambda-1}\!\dots\!
d_{N-1}^{\lambda-1}(1-d_1-\!\dots\!-d_{N-1})^{\lambda-1}.
 \label{Dirichlet}
\end{equation}
where $\lambda > 0$ is a free parameter and $C_{\lambda}$ stands for a
normalization constant. The last component is determined by the trace
condition $d_N=1-d_1-\dots-d_{N-1}$. The uniform measure $\Delta_1$
corresponds to $\lambda=1$. Slater distinguishes also the case $\lambda=1/2$,
which is related to the Fisher information metric \cite{Fisher},
 the Mahalonobis distance \cite{Ma1} and Jeffreys' prior \cite{Jeff},
and was used for many years in different contexts \cite{ECS,Cl,Bal}.
Since this measure is induced by squared elements of a
column (a row) of a random {\sl orthogonal} matrix (see Appendix A), we
shall refer to
\begin{equation}
\mu_o:=\Delta_{1/2} \times \nu_H  \label{muo}
\end{equation}
as to the orthogonal product measure in the space of the mixed quantum
states. Therefore both measures may be directly linked to the well known
Gaussian unitary (orthogonal) ensembles of random matrices \cite{Mehta},
referred to as GUE and GOE.  The measure $\mu_u$ is determined by squared
components of an eigenvector of a GUE matrix, while the measure $\mu_o$  may
defined by components of an eigenvector of GOE matrices \cite{PZK98}. Some
properties of the orthogonal measure $\mu_o$ have been recently studied in
\cite{BBMS98}. Let us stress that the name of the product measure
(orthogonal or unitary) is related to the distribution $\Delta$ on the
simplex $\vec{d}$, while the random rotations $U$ are always assumed to be
distributed according to the Haar measure $\nu_H$ in $U(N)$.

It is interesting to consider the limiting cases of the distribution (\ref
{Dirichlet}). For $\lambda\to 0$ one obtains a singular distribution
concentrated on the pure states only \cite{Sl2}, while in the opposite limit
$\lambda\to \infty$, the distribution is peaked on the maximally mixed state
$\rho_*$ described by the vector ${\vec d}=\{1/N,\dots,1/N\}$. Changing the
continuous parameter $\lambda$ one can thus control the average purity of
the generated mixed states.

\subsection{Separable states}

Consider a composite quantum system described by the density matrix $\rho $
in the $N$ dimensional Hilbert space ${\cal H}={\cal H}_{A}\mathop{\otimes}
{\cal H}_{B}$. Dimension of the system $N$ is equal to the product $%
n_{A}n_{B}$ of the dimensions of the both subsystems. If the state $\rho \in
{\cal H}$ can be expressed as $\rho =\rho _{A}\mathop{\otimes}\rho _{B}$,
with $\rho _{A}\in {\cal H}_{A}$ and $\rho _{B}\in {\cal H}_{B}$, it is
called {\sl product} state (or factorizable state). This occurs if and only
if $\rho ={\rm Tr}_{B}\rho \mathop{\otimes} {\rm Tr}_{A}\rho $, where Tr$_{A}
$ and Tr$_{B}$ denote the operations of partial tracing. In other words, for
such states the description of the composite state is equivalent to the
description in the both subsystems.

A given quantum state $\rho$ is called {\sl separable}, if it can be
represented by a sum of product states \cite{We89}
\begin{equation}
\varrho=\sum_{i=1}^kp_i\varrho_{Ai}\otimes\tilde \varrho_{Bi} ,  \label{sep}
\end{equation}
where $\varrho_{Ai}$ and $\varrho_{Bi}$ are the states on ${\cal H}_B$ and $%
{\cal H}_B$ respectively. The smallest number $k$ of the product states used
in the above decomposition is called the {\sl cardinality} of the separable
state $\rho$ \cite{STV98}.

In general no explicit necessary and sufficient conditions are known for a
mixed state to be separable. However, Peres found a necessary condition
showing that each separable state has the positive partial transpose \cite
{Pe96}. Later Horodeccy demonstrated that for $N=4$ and $N=6$ this is also a
sufficient condition \cite{Ho1}. To represent any state $\rho$ it is
convenient to use an arbitrary orthonormal product basis $|e_j\rangle
\otimes |e_l\rangle $, $j=1,\dots, n_A;$ $l=1,\dots, n_B$ and to define
the
matrix $\rho_{jl,j^{\prime}l^{\prime}}= \langle e_j|\otimes \langle e_l|
\rho |e_j^{\prime}\rangle \otimes |e_l^{\prime}\rangle$. The operation of
{\sl partial transposition} is then defined \cite{Pe96}
\begin{equation}
\varrho^{T_2}_{j l,j^{\prime}l^{\prime}}\equiv \varrho_{j
l^{\prime},j^{\prime}l}.  \label{transpose}
\end{equation}
Even though the matrix $\rho^{T_2}$ depends on the particular based used,
its eigenvalues $\{d^{\prime}_1\ge d^{\prime}_2\ge, \dots \ge d^{\prime}_N\}$
do not. The matrix $\rho^{T_2}$ is positive iff all eigenvalues $d^{\prime}_i
$ are not negative. The practical application of the partial transpose
criterion is thus straightforward: for a given state $\rho$ one computes $%
\rho^{T_2}$, diagonalizes it, and checks the signs of all eigenvalues. To
characterize quantitatively the violation of positivity we introduced \cite
{ZHSL} the negativity
\begin{equation}
t:= \sum_{i=1}^N |d^{\prime}_i| -1,  \label{negat}
\end{equation}
which is equal to zero for all the states with positive partial transpose.

\subsection{Relative volume in the space of the density matrices}

In \cite{ZHSL} we presented several analytical lower and upper bounds for
the volume of separable states. They were obtained assuming the unitary
product measure, but the same reasoning can be repeated for other measures.
The key result: an analytical proof that the volume of separable states is
positive and less than one, is obviously valid for any nonsingular measure.

To analyze the influence of the measure chosen for the volume of the
separable states $P_s$ we picked several random density matrices (circa $10^6
$) distributed according to the orthogonal and unitary product measures, and
verified, whether their partial transpose (\ref{transpose}) are positive.
The results are displayed in Fig.1 as a function of the system size $N$.
Note that for $N>6$ we obtained in this way the volume $P_T$ of the states
with positive partial transposition, which gives an  upper bound for the
volume of separable states. In fact $P_T=P_S+P_B$, where the volume
$P_B$ of the entangled states with
positive partial transpose is studied in the section IV.

The symbols are labeled by the size of the first subsystem $n_A$. For both
measures the symbols seem to lay on one curve, which would imply that
$P_T(n_A,n_B)=P_T(n_A*n_B)$. However, this relation is only approximate,
since $P_T(2\times 6)\ne P_T(3\times 4)$ as pointed out by Smolin \cite{Smo}%
. Numerical results for $P_T$ and $\langle t \rangle$ for $N\le 12$ are
collected in Table 1. The difference between $P_T(2\times 6)$ and $%
P_T(3\times 4)$ is not large, and was smaller than the statistical error of
the results reported in \cite{ZHSL}. Therefore it is reasonable to
neglect for a
while these subtle effects, depending on the way the $N$ dimensional system
is composed, and to ask, how, in a first approximation, $P_T$ changes
with $N$.

\begin{figure}
  \hspace*{-0.3cm}
 \epsfxsize=8.6cm
\epsfbox{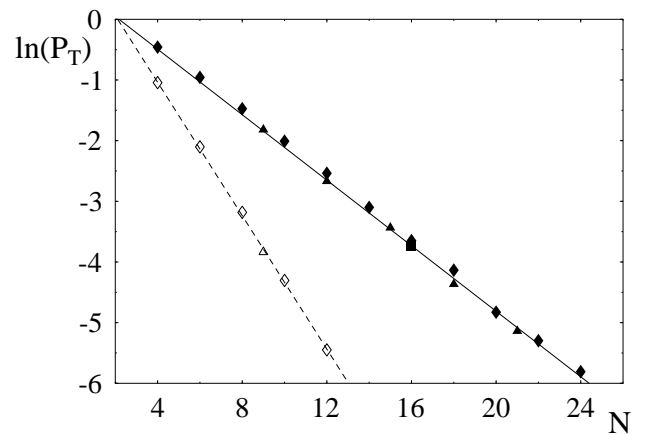} \\
\caption{Probability $P_T$ of finding a state with positive partial
transpose as a function of the dimension of the problem $N$ for unitary
product measure (open symbols) and for orthogonal product measure (full
symbols). For $N\le 6$ it equals to the probability $P_S$ of finding
separable state, while for $N>6$ it gives an upper bound for this quantity.
Different symbols distinguish different sizes of one subsystem; $n_A=$ $%
2(\diamond)$, $3 (\triangle)$ and $4 (\Box )$.}
\label{fig1}
\end{figure}

 \vskip 0.4cm {\bf Table 1.} Probability $P_T$ of
finding a mixed states of
size $N$ with positive partial transpose and the mean negativity
$\langle t \rangle$ for two
product measures orthogonal $\mu_o$ and unitary $\mu_u$. For $N=4$ and
$N=6$
one has $P_T=P_S$.
   
\vskip 0.3cm
\begin{tabular}{|c|c|c||c|c||c|c|}
\hline\hline
~$N$~~ & ~$n_A$~ & ~ $n_B$~ & ~ $\langle P_T\rangle_{\mu_u}$~ & ~~$\langle
t
\rangle_{\mu_u}$~~ & ~ $\langle P_T\rangle_{\mu_o}$~~ & ~~$\langle t
\rangle_{\mu_o}$~ \\ \hline\hline
4 & 2 & 2 & 0.632 & 0.057 & 0.352 & 0.142 \\
6 & 2 & 3 & 0.384 & 0.076 & 0.122 & 0.182 \\
8 & 2 & 4 & 0.229 & 0.082 & 0.042 & 0.204 \\
9 & 3 & 3 & 0.166 & 0.094 & 0.022 & 0.238 \\
10 & 2 & 5 & 0.134 & 0.097 & 0.013 & 0.217 \\
12 & 2 & 6 & 0.079 & 0.098 & 0.0043 & 0.226 \\
12 & 3 & 4 & 0.071 & 0.098 & 0.0039 & 0.266 \\
\hline\hline
\end{tabular}
\vskip 0.4cm 

Figure 1, produced in a semilogarithmical scale, shows that for both
measures the probability $P_T$ decreases
exponentially with the system size $N$.
Obtained numerical results allow us to conjecture
that  lim$_{N\to \infty} P_T(N)=0$
for any (non-singular) probability measure used.
 Observe different slopes of both
lines received for different probability measures. The best fit gives $%
P_{Tu}\sim 1.8 e^{-0.26N}$ for the unitary product measure $\mu_u$, and $%
P_{To}\sim 3.0 e^{-0.55N}$ for the orthogonal product measure $\mu_o$.
Dependence of the probability $P_T$ on the measure chosen is due to the fact
that each measure distinguishes  states of a different purity. This issue
is discussed in details in the following sections.

\section{$2\times 2$ case: positive partial transpose assures separability}

\subsection{Purity versus separability}

For the $N=4$ case the partial transpose criterion is sufficient to assure
the separability\cite{Ho1}, so $P_B=0$ and $P_S=P_T$. Let us investigate,
how the probability of drawing a separable state changes with its purity,
which may be characterized by the von Neumann entropy $H_1(\varrho)=-{\rm Tr}%
(\varrho \ln \varrho)$.  Another quantity, called the participation ratio
\begin{equation}
R(\varrho)={\frac{1}{{\rm {Tr}( \varrho^2)}}},
\end{equation}
is often more convenient for calculations. It varies from unity (for
pure states) to $N$ (for the totally mixed state $\rho_*$ proportional to
the identity matrix) and may be interpreted as an effective number of the
states in the mixture. This quantity gives an lower bound for the rank
$r$ of the matrix $\rho$, namely $r \ge R$. Moreover, it is related to
the von Neumann-Renyi
entropy of order two, $H_2(\varrho)=\ln R(\varrho)$. The latter, also called
the purity of the state, together with other quantum Renyi entropies
\begin{equation}
H_q(\varrho)={\frac{1 }{1-q}} \ln [$Tr$ \varrho^q]  \label{hq}
\end{equation}
is used, for $q\ne 1$, as a measure of how much a given state is mixed (see
e.g. \cite{alfa}).

Figure 2 presents the probability distributions $P(R)$ for $N=4$ density
matrices generated according to the both product measures. As discussed
before, the orthogonal measure $\mu _{o}$ is concentrated at less mixed
states (lower values of $R$) then the unitary measure $\mu _{u}$. For
example the mean value averaged over the orthogonal product measure $\langle
R\rangle _{o}\approx 2.184$ is much smaller than the corresponding mean with
respect to the unitary measure $\langle R\rangle _{u}\approx 2.653$. Observe
a non-smooth behavior of both distributions at $R=3$ ($R=2$),
for which the manifolds of a constant  $R$
start to touch the faces (edges) of the 3D
simplex formed by $d_{1},d_{2}$ and $d_{3}$.

\vskip -1.4cm
\begin{figure}
\hspace*{-1.6cm}
\vspace*{-1.6cm}
\epsfxsize=10.5cm
\epsfbox{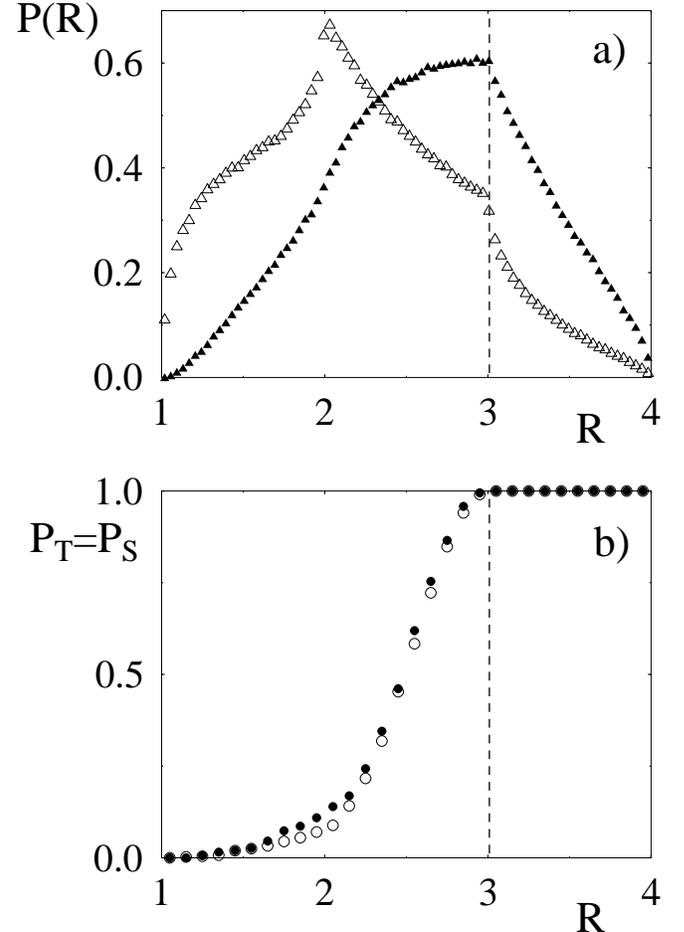} \\
\caption{Purity and separability in $N$=4--dimensional Hilbert space; open
symbols represent averaging over the orthogonal product measure $\protect\mu%
_o$ while closed symbols are obtained with the unitary measure $\protect\mu_u
$; a) probability distributions $P(R)$, b) conditional probability of
finding a separable state as a function of the participation ratio $R$. All
states beyond the dashed vertical line placed at $R=N-1=3$ are
separable.}
\label{fig2}
\end{figure}

Although the distributions $P(R)$ differ considerably for both measures, the
conditional probability of encountering the separable state $P_S(R)$ is
almost measure independent, as shown in Fig. 2b. This is the main result of
this section: the different results obtained for the probability $P_S$ using
various product measures $\mu_{\lambda}$ are due to the different weights
attributed to the mixed states. Since the average mixture $\langle
R\rangle_{\mu_{\lambda}}$ grows monotonically with the parameter $\lambda$
(from $1$ for $\lambda \to 0$ to $4$ for $\lambda\to\infty$)  also the
probability $P_S$ increases with  this parameter from zero to unity. Note
that for both curves the probability $P_S$ achieves unity at $R=3$: all
sufficiently mixed states are separable. This fact has already been
proved in \cite{ZHSL},
 but see also \cite{BCJLPS} for complementary, constructive results.

Above considerations allow us to sketch
the set of entangled states in the case $N=4$.
In the analogy to the Bloch sphere, corresponding to $N=2$,
we take the liberty to depict the set of all quantum states by a ball.
Since it is hardly possible
to draw a picture precisely representing the
complex structure of the $15$ dimensional space of the density matrices,
Fig. 3  should be treated with a pinch of salt. In particular,
the structure of the set of density matrices is not as simple
and  there exist several points inside the ball which do not correspond
to density matrices. Furthermore, the $6$-dimensional space of the pure
states possesses the structure of the complex projective space $CP^3$,
much more complicated than a hypersphere.
 In the sense of the  Hilbert - Schmidt metric,
($\Delta_{HS}(\rho_1,\rho_2)=\sqrt{{\rm Tr}[(\rho_1-\rho_2)^2]}$),
the set of the pure states forms a $6$-dimensional subset of
 the  $14$-dimensional
hypersphere of a radius $\sqrt{3}/2$ centered at $\rho_*={\bf I}/4$.
Keeping this fact in mind, we represent this manifold by a circle
in our oversimplified two dimensional sketch.

\vskip -3.0cm

\begin{figure}
%\hspace*{-1.6cm}
%\vspace*{-1.6cm}
%\epsfxsize=10.5cm
  \hspace*{-0.6cm}
  \vspace*{-1.3cm}
 \epsfxsize=8.6cm
\epsfbox{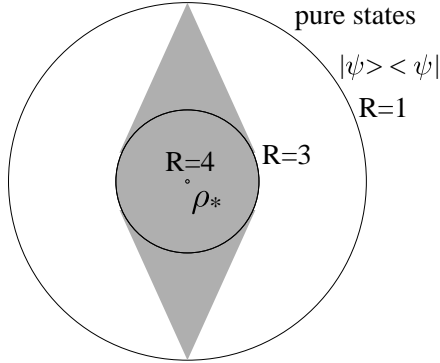} \\
 \vspace*{-3.5cm}
\caption{Sketch of the set of mixed quantum states for $N=4$:
 gray color represents the separable states}
\label{fig3}
\end{figure}

 The set of the separable states is visualized in Fig. 3
as a 'needle of a compass':
it is convex, has a positive measure and includes the vicinity of
the maximally mixed state $\rho_*$. Moreover, it touches the
 manifold of pure states
(pure separable states do exist), but the measure of
this common set is equal to zero.
 The more mixed state,
(localized closer to the center of the 'ball'),
 the larger probability of encountering a separable state.
All states with $R\ge 3$ are separable; for simplicity this complicated 
set is symbolically represented by a smaller circle.

\subsection{Entanglement of formation}

After discussing the problem, how the probability of encountering a
separable state changes with the degree of mixing $R$, we may discuss a
related issue, how the average entanglement depends on $R$. For this purpose
we need a quantitative measure of entanglement of a given mixed state.
Several such quantities have recently been proposed and analyzed \cite
{BVSW96,BBPS96,VPRK97,PR97,VPJK97,VP98,LS98,EP98,Vi98,Ra98,WT98,OK98},
and none of them can be considered as the unique, canonical measure.
However, the quantity called {\sl entanglement of formation}
\cite{BVSW96} plays an important role due to a simple
interpretation: it gives a minimal amount of entanglement necessary to
create a given density matrix.

 For a pure state $|\psi\rangle$ one defines  the von Neuman entropy of
the
reduced state
\begin{equation}
E(\psi)= -{\rm Tr}\rho_A \ln \rho_A = -{\rm Tr}\rho_B \ln \rho_B,
\label{neum}
\end{equation}
where $\rho_A$ is the partial trace of $|\psi\rangle\langle\psi|$ over the
subsystem $B$, while $\rho_B$ has the analogous meaning. This quantity
vanish for a product state. The entanglement of formation of the mixed state
$\rho$ is then defined \cite{BVSW96}
\begin{equation}
E(\rho)= {\rm min} \sum_{i=1}^k p_i E(\Psi_i)  \label{formation}
\end{equation}
and the minimum is taken over all possible decompositions of the mixed state
$\rho$ into pure states
\begin{equation}
\rho= \sum_{i=1}^k p_i |\Psi_i\rangle \langle\Psi_i| ,~ ~ \sum_{i=1}^k p_i
=1.
\end{equation}
The decomposition of $\rho$ into the smallest possible number of $k$ pure
states, for which this minimum is achieved, will be called {\sl optimal
decomposition}, while the number $k$ will be called the {\sl cardinality} of
an entangled state. This definition may be considered as an extension of the
concept of the cardinality of separable states introduced in \cite{STV98},
since for any separable state $\rho_S$ one has $E(\rho_S)=0$.

In Appendix B we present an algorithm allowing one to perform the
minimization crucial for the definition (\ref{formation}). It gives  an
upper estimation of the entanglement of formation for an arbitrary density
matrix of size $N$. The algorithm proposed works fine for $N$ of the order
of $10$ or smaller. In the case of two qubits, discussed in this section, an
analytical solution was found by Hill and Wooters \cite{HW97,Wo98}, who
introduced the concept of {\sl concurrence}.

For any $4\times 4$ density matrix $\rho$ one defines the flipped state $%
\tilde{\rho}=O\rho^*O^{T}$, where $\rho^*$ denotes the complex conjugation
and the orthogonal flipping matrix $O$ contains only four nonzero elements
along the antidiagonal: $O_{14}=O_{41}=1$ and $O_{23}=O_{32}=-1$.
Concurrence $C(\rho)$ is then defined \cite{HW97}
\begin{equation}
C(\rho):= {\rm max} \{0,\alpha_1-\alpha_2-\alpha_3-\alpha_4 \},
\label{concur}
\end{equation}
where $\alpha_i$s are the eigenvalues, in decreasing order, of the
Hermitian matrix $\sqrt{\sqrt{\rho} {\tilde{\rho}}\sqrt{\rho}}$.
Note that this matrix determines the Bures distance \cite{Ul76}
 between $\rho$ and $\tilde{\rho}$. In other
words $\alpha_i$s are the non-negative square roots of the moduli of the
complex eigenvalues of the non Hermitian matrix $\rho {\tilde \rho}$.

Concurrence $C$ of a given state $\rho$ determines its entanglement of
formation \cite{HW97,Wo98}
\begin{equation}
E(\rho) = h\Bigr( {\frac{1 }{2}} [1+\sqrt{1-C^2(\rho)}]\Bigl),
\label{conc2}
\end{equation}
where
\begin{equation}
h(x) := -x \ln(x) - (1-x)\ln(1-x)
\end{equation}
is the Shannon entropy of the $2$-elements partition $\{x,1-x\}$.

\vskip -1.85cm
\begin{figure}
\hspace*{-1.8cm}
\vspace*{-1.45cm}
 \epsfxsize=10.6cm
 \epsfbox{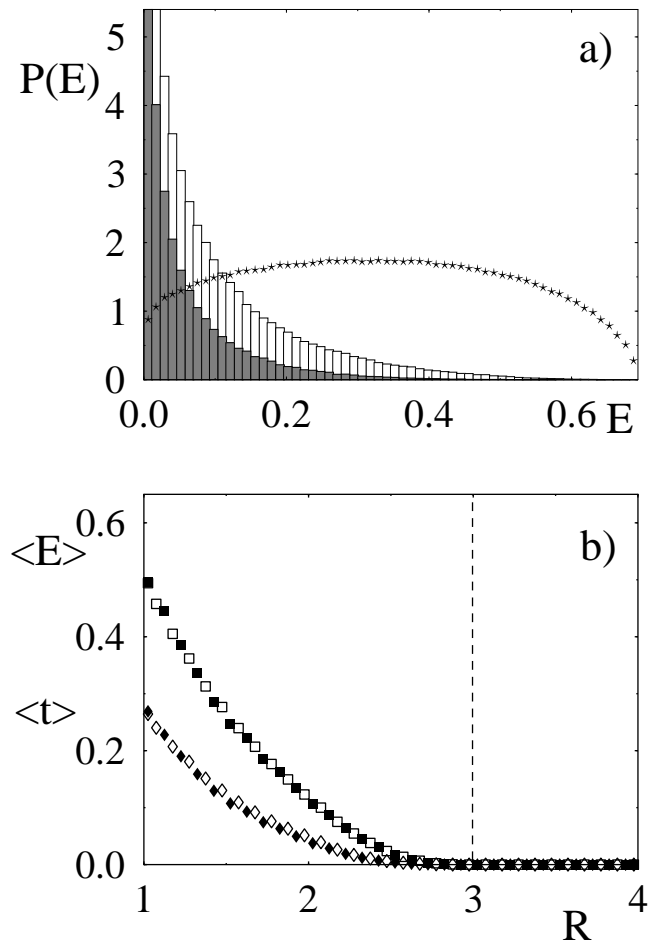} \\
\caption{$2\times 2$ system; a) The distributions $P(E)$ of the entanglement
of formation obtained for the density matrices generated according to $%
\protect\mu_o$ (white histogram, open symbols) and $\protect\mu_u$ (gray
histogram, closed symbols), and
the rotationally uniform distribution in the set of pure states $(*)$; b)
average entanglement $E(R)$ (squares) and average negativity $t(R)$
(diamonds) for both measures $\mu$.}
\label{fig4}
\end{figure}

Note that in the definition of entropy (\ref{neum}) the natural logarithm
was used, (in contrast to the binary logarithm present in the paper \cite
{HW97}), so the entanglement $E \in [0, \ln 2]$. Two histograms in
Figure~4
present the probability distribution $P(E)$ obtained for $N=4$ random
density matrices distributed according to both product measures $\mu_o$ and $%
\mu_u$. The singular peak at $E=0$, corresponding to the separable states,
is omitted. Large entanglements of formation are rather unlikely. The mean
values are not large: $\langle E\rangle_o\approx 0.055$ and $\langle
E\rangle_u\approx 0.018$, since the averages are influenced by a
considerable fraction of separable states with $E=0$. The probability of
obtaining a given value of $E$ is larger for the orthogonal measure, which
gives favor to more pure, and more likely entangled states.

Both histograms may be compared with the probability distribution $P(E)$
obtained for the ensemble of pure states, represented by stars in Fig. 4a.
This distribution is less peaked; vaguely speaking, different degrees of
entanglement are almost equally likely among the pure states. The minimum of
probability can be observed for maximally entangled states ($E=\ln 2$),
while the mean $\langle E \rangle_{{\rm {pure}}}\approx 0.328$ is close to $%
(\ln 2)/2$. Since the singular distribution concentrated exclusively on pure
states corresponds to the case $\lambda\to 0$ in the distribution (\ref
{Dirichlet}),  we observe that the mean entanglement $\langle E \rangle $
decreases with increase of the parameter $\lambda$, as the distributions
$\Delta_{\lambda}$ increasingly favor more mixed states.

Although the mean entanglement $\langle E \rangle$ strongly depends on the
measure used, the conditional mean entanglement $E(R)$, averaged over all
states of the same degree of mixing $R$, is not sensitive to the choice of a
measure, as demonstrated in figure 4b. This allows us to formulate a
general
quantitative conclusion, valid for nonsingular measures in the space of
density matrices: {\sl the larger the average degree of mixing $R$, the smaller
the  mean
entanglement of formation $E$}. For $R > 3 $ one has $E(R)=0$ \cite{ZHSL}.

\subsection{Negativity and concurrence}

In Ref. \cite{ZHSL} we proposed a simple quantity $t$ defined by (\ref{negat}%
), which characterizes quantitatively, to what extent the positivity of
partial transpose is violated. As shown in Fig. 4b the conditional average
$t(R)$ does not depend on the measure applied and decreases monotonically
with $R$. This dependence resembles the function $E(R)$, which suggests a
possible link between the both quantities.

To analyze such a relation between these measures of entanglement, following
the strategy of Eisert and Plenio \cite{EP98}, we generated $10^5$ random
density matrices $\rho$ computing their concurrence $C$, entanglement $E=E(C)
$ and negativity $t$. As expected, the points at the plot $E$ versus $t$ do
not form a single curve. It means that both quantities, entanglement of
formation and the negativity, {\sl do not} generate the same ordering in the
space of $4\times 4$ density matrices. However, large correlation
coefficients (approximately $0.978$ for the orthogonal measure and $0.967$
for the unitary measure) reveals a statistical connection between these
measures.

\begin{figure}
\hspace*{-0.3cm}
\vspace*{-0.3cm}
 \epsfxsize=8.5cm
 \epsfbox{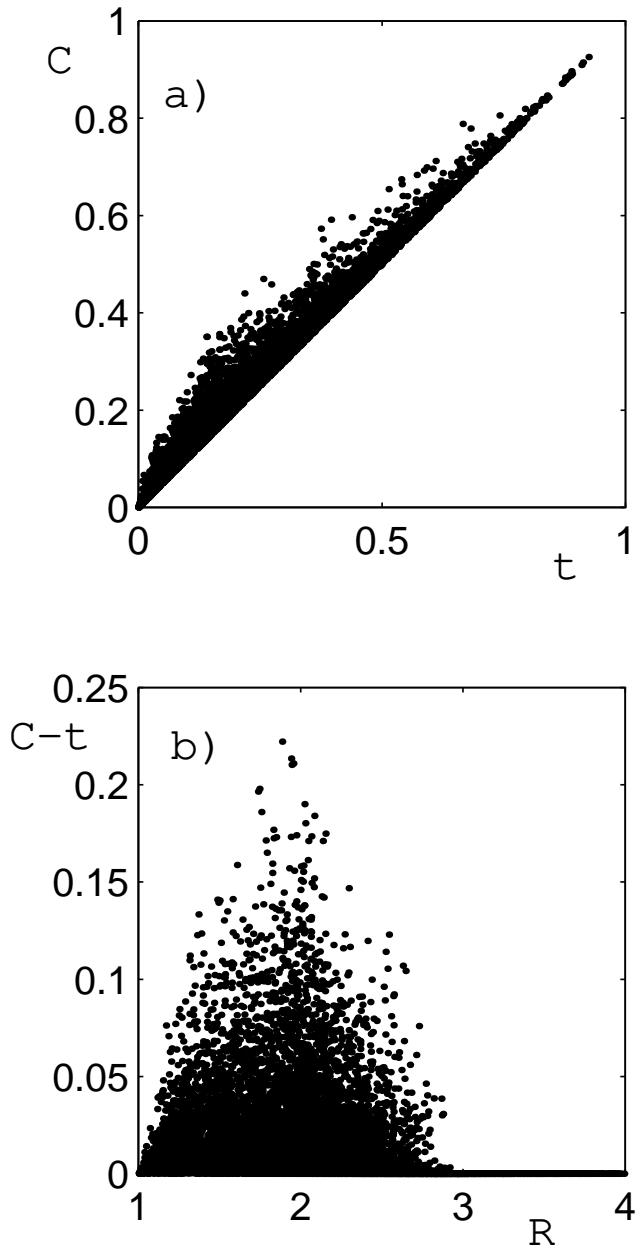} \\
\caption{Ten thousands random density matrices of the size $N=4$ distributed
according to the orthogonal product measure; a) plot in the plane negativity
-- concurrence; b) plot of the difference $C-t$ versus the participation $R$.}
\label{fig5}
\end{figure}

It is particularly useful to look at the plane concurrence versus
negativity. The data presented in Fig. 5a are obtained with the measure $%
\mu_o$. We observed, independently of the measure used, that all points are
localized at or above the diagonal. This allows us to conjecture  that for
any density matrix $\rho$ the following inequality holds
\begin{equation}
t(\rho) \le C(\rho).  \label{conj}
\end{equation}
A similar observation was already reported in \cite{EP98}, where a modulus
of the negative eigenvalue $E_N$ of the partially transposed matrix $%
\rho^{T_2}$ was used. Since for $N=4$ no more than one eigenvalue $%
d^{\prime}_4$ is negative \cite{STV98}, both quantities are equivalent and $%
t=2E_N$. Note that due to the conjecture (\ref{conj}) we can attribute to
negativity a more specific meaning. By means of equation (\ref{conc2})
and the fact that $h(x)$ decreases for $x>1/2$, negativity $t$
allows us to obtain a lower bound for the entanglement of formation $E$.

Numerical investigations show that the difference $C-t$ is largest for mixed
states with $R\approx 2$, while it vanish for $R\ge 3$ and $R=1$ (see Fig.
5b). In the former case all states are separable and $C=t=0$. The latter
case corresponds to pure states for which $\alpha_2=\alpha_3=\alpha_4=0$
\cite{Wo98} and $C=\alpha_1=-2d^{\prime}_4=t$. Thus the inequality (\ref
{conj}) becomes sharp for separable states or pure states.

\subsection{Mixed states with the same partition ratio $R$}

As demonstrated in Fig.2b the conditional probability $P_S(R)$ of
encountering a
separable state is similar for states with the same participation $R$,
averaged over both product measures $\mu_o$ and $\mu_u$. This does not mean,
however, that the probability $P_S$ is constant for each family of states $%
\rho=U d U^{\dagger}$ defined by a given vector $d$ with fixed
participation
ratio $R$. To illustrate this issue we discuss the case $R=2$.

Consider a vector of eigenvalues $\vec{d}$ with $r$ nonzero elements.
This  natural
number ($r\in [1,4]$) is just the rank of the matrix $\rho$.
Any state $\rho=U d U^{\dagger}$ can
be expressed by the sum of $r$ terms, $\rho_{ij}=\sum_{l=1}^r d_l
U_{il}U^*_{jl}$.  Moreover, the number of nonzero eigenvalues $\alpha_i$
entering the definition of concurrence (\ref{concur}) equals to $r$
\cite  {Wo98}. 

Take any vector with $r=2$ nonzero elements. In this case the formula
(\ref
{concur}) reduces to $C=\alpha_1-\alpha_2$. Since per definition $%
\alpha_1\ge\alpha_2$, the concurrence is positive unless $\alpha_1=\alpha_2$%
. Such degenerate cases occur with probability zero, (e.g. for diagonal
rotation matrices $U$), so one arrives with a simple conclusion:

For any set $d$ of eigenvalues with $r \le 2$ the probability
$P_S$ that a random state $UdU^{\dagger}$ is separable, is equal to
zero.

For concreteness consider three vectors of eigenvalues
characterized by $r=2,3$ and $4$. We put
 $\vec{d_a} =\{1/2,1/2,0,0 \}$,
 $\vec{d_b}=\{2/3,1/6,1/6,0 \}$ and $\vec{d_c}%
=\{x_1,x_2,x_2,x_2 \}$, where $x_1=(1+\sqrt{3})/4$ and
 $x_2=(1-x_1)/3$. Each
such vector generates an ensemble of density matrices
$\rho=UdU^{\dagger}$,
where $U$ stands for a random unitary rotation matrix of the size $N=4$.
Although all three ensembles are characterized by the same participation
ratio $R=2$, the probabilities of generating an separable state are
different. The case $\vec{d_a}$ is characterized by $r=2$, so $P_S=0$.
Numerical results obtained of a sample of $10^5$ random unitary matrices
give $P_S\approx 0.105$ and $0.200$ for $\vec{d_b}$ and $\vec{d_c}$,
respectively. Thus the probability $P_S$ grows with the number $r$ of
pure
states necessary to construct given mixed state $\rho$, or with the von
Neuman entropy $H_1$.

On the other hand, the average quantities characterizing entanglement
(negativity, concurrence or entanglement of formation) decrease with
$r$,
provided, the participation $R$ is fixed. For example, the mean
entanglement, $\langle E \rangle$,
equals $0.063$, $0.057$ and $0.042$ for the ensembles $\vec{d_a}$,
$\vec{d_b}$ and $\vec{d_c}$, respectively. Interestingly, in the latter case
(or any other ensemble with $d_2=d_3=d_4$), one has $\alpha_3=\alpha_4$ and $%
C=t$.

\section{$2\times 4$ case: Positive partial transpose does not assure
separability}

\subsection{Purity and positive partial transpose}

For any system size the probability of finding the states with positive
partial transpose depends on the measure used, as shown in Fig.1 and Table
1. On the other hand, for any $N$ the relations between purity and
entanglement depend only weakly on the kind of the product measure used.

\vskip -1.6cm
\begin{figure}
\hspace*{-1.7cm}
\vspace*{-1.3cm}
 \epsfxsize=10.0cm
 \epsfbox{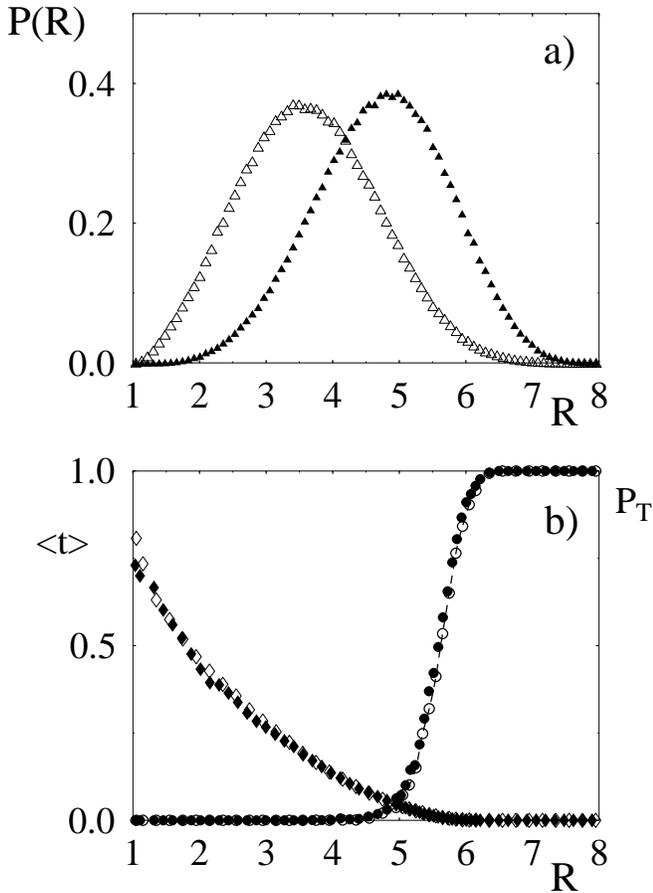} \\
\caption{As in Fig.2 for the $2\times 4$ system $(N=8)$. The circles in Fig.
6b represent the conditional probability of finding a state with positive
partial transpose, $P_T(R)$. Diamonds represent the average negativity $t(R)$
obtained with the measures $\protect\mu_o$ (open symbols) and $\protect\mu_u$
(full symbols).}
\label{fig6}
\end{figure}

Figure 6b presents the conditional probability of $P_T(R)$ and the
mean negativity $t$
as a function of the degree if mixing $R$ for the
$2\times 4$
system. For both quantities the results obtained with orthogonal and unitary
product measures are difficult to distinguish. Thus the dependence of the
total probability $P_T$ on the measure used is strongly influenced by the
likelihood of generating highly mixed states, described by the distribution $%
P(R)$.

These distributions for $N=8$ are shown in Fig. 6a. The histogram for the
unitary measure $\mu _{u}$ is shifted to larger values of $R$, with respect
to the data obtained with the orthogonal measure. Quantitatively, the mean
values read $\langle R\rangle _{u}\approx 4.74>\langle R\rangle _{o}\approx
3.66$. It is known that $P_{T}=1$ for $N>R-1$ \cite{ZHSL}. The right
histogram, corresponding to the unitary measure, has a larger overlap with
this region, which causes $\langle P_{T}\rangle _{u}>\langle P_{T}\rangle
_{o}$.

This observation is valid for an arbitrary matrix size, since for large $N$
one has $\langle R (N)\rangle_u \approx N/2$ while $\langle R (N)
\rangle_o\approx N/3$ \cite{ZK94}. For $N$ large enough the
distributions $P(R)$ tend to Gaussians. They are centered at the mean
values, which do depend on the measure, while the variance $\sigma^2$ is of
the order of $N/5$ for both measures under the consideration.
Hence, the overlap with the interval $[N-1,N]$ is larger for the
measure $\mu_u$ characterized by a larger mean value $\langle R
\rangle_u$.

\subsection{Entanglement of formation}

Since for $N>4$ there exist no analytical methods to compute the
entanglement of formation of an arbitrary mixed state $\rho$, we relied on
numerical computations. To perform the minimization present in the
definition (\ref{formation}) we worked out an algorithm based on a random
walk in the space of unitary matrices $U(M)$ with $M\ge N$. It is
described in detail in
Appendix B. Each run ends with an approximate {\sl optimal decomposition} of
the state $\rho$ and provides an {\sl upper} estimation of the entanglement $%
E$. To verify the accuracy of this technique we started with for the case $%
N=4$, in which the explicit formula (\ref{conc2}) is known. Computing
numerically entanglement for $1000$ randomly chosen $N=4$ mixed states we
obtained the mean error of the order of $10^{-7}$, while the maximal error
was smaller than $10^{-4}$.

At the beginning of each computation one has to choose the number $M$,
determining the number of pure states in the decomposition. Since for $N=4$
it is known that the cardinality of any state is not larger than $4$ \cite
{Wo98,STV98}, it is sufficient to look
 for the optimal decomposition in the
$M=N=k=4$ dimensional space. For larger systems the problem of finding the
maximal possible cardinality is open. For each randomly generated mixed
state $\rho$
in the discussed  $2\times 4$ case we started to look for the optimal
decomposition with $M=N=8$, recorded the minimal entanglement $E_{M=8}$, and
repeated computations with $M=9,10,\dots,M_{max}$. It is known
\cite{Ul97,Ho2} that
the maximal number of pure states does not exceed $N^2$,
but in practice we analyzed $M\in [N,2N]$

The number of degrees of freedom grows as $M^2$, so the process of
search
for the optimal decomposition becomes less efficient with increase of the
number $M$. However, for certain states we found better estimations for
entanglement, e.g. $E_{M=9}(\rho) < E_{M=8}(\rho)$. In these rare cases, the
improvements of the estimations of $E$ were very small, and repeating
several
times our procedure with $M=8$ the same upper bounds for entanglement of
formation were reproduced.

Thus our results do not contradict an appealing conjecture
that the
{\sl cardinality $k$ of any $2 \times 4$
mixed system is not larger than $N=8$}.
Further work is still needed to verify, whether this conjecture is true.

Note that the numerical algorithm to search for the optimal decomposition
and the entropy of formation, may also be used to look for the generalized
entropy of formation $E_q$,  in the analogy to (\ref{hq}) and (\ref
{formation}), see \cite{Vi98}. We found it interesting to study the quantity
$E_2$, which has a similar interpretation as the participation ratio $R$,
and equals to unity for the separable states.

\subsection{Volume of the bound entangled states}

It is known \cite{Ho3} that
for $N=8$ there exist bound entangled states, which cannot be brought
into the singlet form. All entangled states satisfying
the partial transposition criterion are bound entangled
\cite{Ho2,Ho3,VMSST}. It was shown in \cite{ZHSL}
that they occupy a positive volume $P_B$. Therefore $P_S=P_T-P_B$ is
smaller than the volume $P_T$ of the states with positive partial
transpose. Strictly speaking, the volume $P_B$
 of entangled states with positive partial transpose
should be considered as a lower bound of the volume of bound entangled
states, since it is not proven yet that all states
with negative partial transpose are free entangled.

To estimate $P_B$ we generated $10^5$ random density matrices of size $N=8$.
We worked with the unitary product measure $\mu_u$, since, as shown in
Table~1, the $2\times 4$ states chosen according to the orthogonal measure $%
\mu_o$ very seldom satisfy the partial transposition criterion. To save the
computing time we estimated the entanglement of formation $E$ only in the $%
2223$ cases with positive partial transpose. Setting an entanglement cut-off
$E_c=0.0003$, (see appendix B), we found that $473$ states enjoyed the
entanglement $E>E_c$. This gives a fraction of $P_B\approx 4.7\%$ of all
states, or $P_B/P_T=21.3\%$ of the states with positive partial transpose.
Although these numbers are influenced by systematic errors (bound
entangled states with $E<E_c$ are regarded as separable, while separable
states with numerically obtained upper estimations of the
entanglement larger than $E_c$ are
considered as entangled), the dependence
of the results obtained on the
cut-off value $E_c$ is weak. Moreover, these results do not depend on
the exact values of the parameters characterizing the random walk (see
Appendix B). Consequently, we obtained an estimate of the
volume of separable states for this case, $P_S=P_T-P_B\approx 17.5\%$,
as shown in the inset of Fig.7.

\subsection{Bound entanglement and  purity}

It is interesting to ask, whether a certain degree of mixing favors the
probability of finding the bound entangled states. Grouping all $10^{5}$
analyzed states in $30$ bins according to the participation ratio $R$,
we computed
the conditional probabilities of entanglement. These results are shown in
Fig. 7. Probability $P_{S}$ increases monotonically with $R$ while the
probability of finding a free entangled state $P_{F}=1-P_{T}$ decays with
the participation. On the other hand, the conditional probability $P_{B}(R)$
of finding a bound entangled state exhibits a clear
 maximum at $R\sim 5.5$. If
the mean purity is concerned, the bound entangled states are thus sandwiched
between free entangled states (generally of high purity) and the separable
states characterized by a high degree of mixing.

  \vskip -7.7cm
\begin{figure}
\hspace*{-2.2cm}
\vspace*{-1.5cm}
 \epsfxsize=10.5cm
 \epsfbox{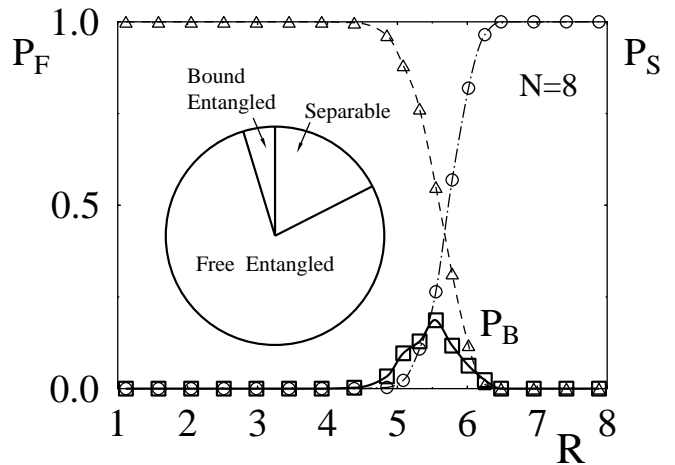} \\
\caption{Conditional probabilities of finding the separable
 states $(\circ)$,
free entangled states $(\triangle)$ and bound entangled states
$(\Box)$ as a
function of the participation ratio $R$. Results are obtained with $10^5$
random density matrices of the size $N=8$ distributed according to the
measure $\protect\mu_u$. The lines are drawn to guide the eye.
The inset shows the pie-chart of the total probability of encountering
separable states,  bound entangled states (lower bound), and
free entangled states (upper bound).}
\label{fig7}
\end{figure}

Above results suggest that for the bound entangled states
 there exist a minimal participation ratio $R$
or a minimal rank $r$. Preparing a sketch analogous to Fig. 3 for
$N=8$, one should put the bound entangled states close to the center
of the figure, but outside the symbolic
'needle of the compass', which  represents the separable states.

It is worth noting, that the entanglement of formation for bound entangled
states is rather small in comparison to the mean entanglement of formation
for the free entangled states, which violate the partial transposition
criterion. The average taken over all free entangled states is $\langle E
\rangle_F\approx 0.05$, the average taken over all bound entangled states is
$\langle E \rangle_B \approx 0.0033$, (which is much larger than the
cut-off value $E_c$). The maximal entanglement found
for a bound state was only $0.0746$.

\section{Closing Remarks}

In this work we provide an attempt to characterize the
statistical properties of the
set of separable mixed quantum states.
Certain level of caution is always recommended by interpretation of any
results of probabilistic calculations, especially if the space of the
outcomes is infinite. Let us mention here the famous
Bertrand's paradox: what is the probability that a randomly chosen chord
of
a circle is longer than the side of the equilateral triangle inscribed
within the circle? The answer depends on the construction of the
randomly chosen chord, which determines the measure in the infinite
space of the possible outcomes.

Asking a question on the probability that a randomly chosen mixed
state is separable, one should also expect that the answer will depend
on the measure used. This is indeed the case, as demonstrated in this
work for two products measures, and also shown by Slater
\cite{Sl2} for a different measure
related to the {\sl monotone} metrics \cite{PS96}.
We reach, therefore, a simple conclusion, rather intuitive for an
experimental physicist:
the probability of finding an entangled state depends on the way, the
states are prepared, which determines
the measure in the space of mixed quantum states.

On the other hand, in this paper we provide arguments supporting
the conjecture that some statistical properties of entangled
states are universal
and to a large extend do not depend (or depend rather weakly)
on the measure used. Let us mention only the exponential decay of the
volume of
the set of the separable states with the size $N$ of the problem or the
important relation between the purity
of mixed quantum states and the probability to find a separable state.

Studying the simplest case $N=4$ we have shown that the distribution of
entanglement of formation is close to uniform in $[0,$ln$2]$ for all
pure states. The more mixed states, the larger peak at small values of
entanglement, the larger probability of finding a separable state. We
have shown that the negativity $t$, a naive measure of entanglement,
provides a lower bound for the entanglement of formation.

Analyzing the more sophisticated problem $N=8$ we developed an efficient
numerical
algorithm to estimate the entanglement of formation of any mixed state.
In this way we could differentiate between separable states and the
bound entangled states.  About $79\%$ of
$N=8$ states satisfying the positive transposition criterion are
separable. This result is obtained for random states generated according to
the unitary product measure in the space of $N=8$ density matrices, but we
expect to get comparable results for other, non-singular measures.
Mean entanglement of formation for the bound entangled states is much
smaller than for the free entangled states.
 Relative probability of finding
a bound entangled state for the $2 \times 4 $ systems is largest for
moderately mixed systems, characterized by the participation ratio close to
$R=5.5$.

Even though this paper follows the previous work \cite{ZHSL}, the list
of open problems in this field is still very long. Let us collect here some
of them related to this work, mentioning also these, already discussed
in the literature.

\smallskip

A. $N=4$, ($2\times2$ systems)

(i)  Check, whether the dependence of the conditional
probability on the participation ratio, $P_S(R)$, obtained for
two product
measures (see Fig. 2b) holds also for the measures based on the monotone
metrics \cite{Sl2} or for the product Bures measure \cite{Sln,Ha98}.

(ii) Prove the relation between the concurrence and the negativity: $C
\ge t$.

(iii) Find max$(C-t)$ as a function of the participation ratio $R$, (see
Fig.~4b).

(iv) Check whether the following conjecture is true: If
$R(\vec{d}_1)=R(\vec{d}_2)$ and $H_1(\vec{d}_1)\ge H_1(\vec{d}_2)$
then $P_S(\rho_1) \ge P_S(\rho_2)$.
The von Neuman entropy $H_1$ and the participation ratio $R$ measure the
degree of mixing of a given vector $\vec{d}$, while $P_S$ denotes the
probability that a random state $\rho_i=U d_i U^{\dagger}$ is separable.

(v) Find iso-probability surfaces in the simplex $\{d_1,d_2,d_3\}$ such
that $P_S(\vec{d})=$const.

\smallskip
B. $N=6$, ($2\times 3$ or $3\times 2$ systems)

(vi) Find a lower bound for the entanglement of formation $E$,  (in the
analogy to negativity $t$,
which gives a lower bound for $C$ and $E$ in the case $N=4$).

(vii) Find an explicit formula for $E$ in this case.

\smallskip
C. $N=8$, ($2\times 4$ or $4\times 2$ systems)

(viii) Find necessary and sufficient conditions for a bound entangled
(or separable) state.

(ix) Find the maximal entanglement of formation $E$ of a bound entangled
state.

(x) Check whether the rank of bound entangled states is bounded from
below.

(xi) Check whether all states violating the partial
     transpose criterion are free entangled.

(xii) Check whether the cardinality of any state is not larger than $8$.

\smallskip
D. General questions

(xiii) Verify, whether the optimal decomposition of a given mixed state
into a
sum of pure states leading to the entanglement of formation, $E=E_1$, also
gives the minimum of the generalized entanglement of formation $E_q$.

(xiv) For what $N_A \times N_B$ composed systems
  the cardinality $k$ of any mixed state in the $N$ dimensional
  Hilbert space  is less or equal to $N=N_AN_B$?

(xv) Check whether the entanglement of formation is additive.

\smallskip
Not all of the above problems are of the same importance. As most relevant
we regard the questions (i),(viii) and the last two general problems.
Preliminary results of Slater suggest \cite{Slano}
that the relation $P_S(R)$ for monotone metrics is similar
to this obtained here for product metric, at least for $N=4$.
Concerning the question (viii): for separable states
with $N=8$, some necessary conditions, stronger than the positive
partial transpose, are known \cite{hopc,LCK99}, but the sufficient
conditions
assuring the separability are still most welcome. The problem of additivity
of entanglement of formation is present in the literature (see e.g. \cite
{Wo98}). Performing numerical estimations of the entanglement of formation $E$
 for several states of  $2 \times N_B$ systems
 we have not found any cases violating the statements (xiv)
 and (xv) \cite{HoZy}.
Recent results of Lewenstein, Cirac and Karnas \cite{LCK99}
suggest that the answer for the problem (xiv) is negative for
the systems $3 \times N_B$ with $N_B > 3$,
but do not contradict this statement for
the $2 \times N_B$ composed systems. Further effort is
required to establish whether in this case the answer for the problem
(xiv) is positive.

\vskip 0.2cm {\bf Acknowledgments}

I am very grateful to P.~Horodecki and P.~Slater for motivating
correspondence, fruitful interaction
and a constant interest in this work. It is
also a pleasure to thank J.~Smolin and W.~S{\l}omczy{\'n}ski for useful
comments  and M.~Lewenstein and A.~Sanpera for
several valuable remarks and the collaboration at the early
stage of this project.
I appreciate the helpful remarks from an anonymous referee.
This work has been supported by the grant no.
P03B~060~013 financed by the Komitet Bada{\'n} Naukowych in Warsaw.

\appendix

\section{Rotationally invariant product measures}

In this appendix we show that a vector of a $N$-dimensional random
orthogonal (unitary) matrix generates the Dirichlet measure (\ref{Dirichlet}%
) with $\lambda =1/2$ ($\lambda =1$) in the $(N-1)$ D simplex. Although
these results seem not to be new, we have not found them in the literature
in this form and prove them here for convenience of the reader, starting
with the simplest case $N=2$.

{\bf Lemma 1}. {\sl Let $O$ be a $N\times N$ random orthogonal matrix
distributed according to the Haar measure on $O(N)$. Then the vector $%
d_{i}=|O_{i1}|^{2};i=1,\dots ,N$ is distributed according to the statistical
measure on the  }${\sl (N-1)}$ {\sl dimensional simplex (Dirichlet measure
with $\lambda =1/2$).}

{\bf Proof.} Due to the rotational invariance of the Haar measure on $O(N)$
the vector $O_{i1}$ is distributed uniformly on the $N-1$ dimensional sphere
$S^{N-1}$. Thus $\sum_{i=1}^N d_i=1$.

For $N=2$ the vector $|O_{i1}|$ is distributed uniformly along the quarter of
the circle of radius $1$. Therefore $x=\cos \phi $, where
$\phi \in [0,\pi/2)$, and $P(\phi)=2/\pi$.
 Hence $P(x)=P(\phi )d\phi /dx=2/(\pi \sqrt{1-x^{2}})$.
Another substitution $\xi =x^{2}$ gives the required result:
$P(\xi )=P(x)dx/d\xi =1/(\pi \sqrt{\xi (1-\xi )})$.

To discuss the general, $N$ dimensional case, it is convenient to introduce
the polar angles and to represent any point belonging to the $(N-1)$ D
sphere as $x_{N}=\cos \theta _{N-2}$, $\rho =\sin \theta _{N-2}$, where
$\rho ^{2}=1-\sum_{i=1}^{N-1}x_{i}^{2}$. Uniform distribution of the points
on the sphere is described by the volume element $d\Omega =\sin ^{N-2}\theta
_{N-2}d\theta _{N-2}\cdots \sin \theta _{1}d\theta _{1}d\phi $. Changing the
polar variables into Cartesian we obtain $P(\rho )\sim 1/\cos \theta
_{N-2}=1/\sqrt{1-\rho ^{2}}$. The last change of the variables $\xi
_{i}:=x_{i}^{2}$ for $i=1,\dots ,N$ allows us to receive $P(\xi _{1},\dots
,\xi _{N-1})\sim $ $[\xi _{1}\xi _{2}\dots \xi _{N-1}(1-\xi _{1}-\xi
_{2}-\cdots -\xi _{N-1})]^{-1/2}$, which gives the statistical measure $%
\Delta _{1/2}$ defined in Eq.~(\ref{Dirichlet}). $\Box $.

Geometric interpretation of this result is particularly convincing
for $N=3$.
Then the vector $O_{i1}$ covers uniformly the sphere $S^{2}$, while
$|O_{i1}|$ is distributed uniformly in the first octant. The points
$\{d_{1},d_{2},d_{3}\}=\{\xi _{1},\xi _{2},\xi _{3}\}$ lay at the plane
$z=1-x-y$. Their projection into the $x-y$ plane gives the statistical
measure on the 2D simplex, i.e. the triangle
$\left\{(0,0),(1,0),(0,1)\right\}$.

{\bf Lemma 2}. {\sl Let $U$ be a $N\times N$ random unitary matrix
distributed according to the Haar measure on $U(N)$. Then the vector
$d_{i}=|U_{i1}|^{2}, i=1,\dots,N$ is distributed according to the
uniform measure on the}
${\sl (N-1)}${\sl \ dimensional simplex
(Dirichlet measure with $\lambda =1$).}

{\bf Proof}. We will use the Hurwitz parametrization of $U(N)$ \cite{PZK98},
based on the angles $\varphi_{kl}\in [0,\pi/2]$ with $0\le k < l \le N-1$.
Their distribution can be determined by the relation $\varphi_{kl}=$arcsin$%
\xi_{k+1}^{1/(2k+2)}$, where $\xi_{k}$ are the auxiliary independent random
numbers distributed uniformly in $[0,1]$ (see Ref. \cite{PZK98}).

In the simplest case $N=2$ the vector $\vec{d}$ reads $|U_{i1}|^{2}=\{\cos
^{2}\varphi _{01},\sin ^{2}\varphi _{01}\}=\{\xi _{1},1-\xi _{1}\}$ and the
variable $d_{1}=\xi _{1}$ is distributed uniformly in the interval $[0,1]$
(one dimensional simplex). For $N=3$ one obtains $\vec{d}=\{\cos ^{2}\varphi
_{12},\sin ^{2}\varphi _{12}\cos ^{2}\varphi _{01},\sin ^{2}\varphi
_{12}\sin ^{2}\varphi _{01}\}=$
$\{1-\xi _{2}^{1/2},\xi_{2}^{1/2}(1-\xi _{1}),\xi _{2}^{1/2}\xi_{1}\}$,
which is distributed uniformly in the
simplex $\left\{ (0,0),(1,0),(0,1)\right\} $.

In the general $N$-dimensional case we get $\vec{d}=\{\cos ^{2}\varphi
_{N-2,N-1},\sin ^{2}\varphi _{N-2,N-1}\cos ^{2}\varphi _{N-3,N-1},$
$\sin ^{2}\varphi _{N-2,N-1}\sin ^{2}\varphi _{N-3,N-1}\cos ^{2}\varphi
_{N-4,N-1},\dots ,$
$\sin ^{2}\varphi _{N-2,N-1}\cdots \sin ^{2}\varphi _{1,N-1}\cos ^{2}\varphi
_{0,N-1},$
$\sin ^{2}\varphi _{N-2,N-1}\cdots \sin ^{2}\varphi _{1,N-1}\sin ^{2}\varphi
_{0,N-1}\}$. Using uniformly distributed random variables this vector may be
written as $\{1-\xi_{N-1}^{1/(N-1)},
\xi_{N-1}^{1/(N-1)}(1-\xi_{N-2}^{1/(N-2)}),$
$\xi_{N-1}^{1/(N-1)}\xi_{N-2}^{1/(N-2)}(1-\xi_{N-3}^{1/(N-3)}),\dots,$
$\xi_{N-1}^{1/(N-1)}\cdots \xi_{2}^{1/2}(1-\xi_{1}),
\xi_{N-1}^{1/(N-1)}\cdots \xi_{2}^{1/2}\xi_{1}\}$.
 This vector is uniformly
distributed in the $N-1$ dimensional simplex, as explicitly shown in
Appendix A of \cite{ZHSL}. $\Box $.

The above lemmae allow one to generate random points
distributed in the simplex according to the both measures using vectors of
random orthogonal (unitary) matrices. They may be constructed
according to the
algorithms presented in Ref. \cite{PZK98}. Alternatively, one may take a
random matrix of Gaussian orthogonal (unitary) ensemble, diagonalize it, and
use one of its eigenvectors as in (\ref{dii}). Random matrices pertaining to
GOE (GUE) are obtained as symmetric (Hermitian) matrices with all elements
given by independent random Gaussian variables. Several ensembles
interpolating between GOE and GUE are known \cite{Mehta}. Statistics of
eigenvectors during such a transition were studied e.g. in \cite{ZL91};
while the
transitions between circular ensembles of unitary matrices were analyzed in
\cite{ZK96}.

\section{Entanglement of formation - a numerical algorithm}

\subsection{Generating random density matrix}

In oder to generate an $N \times N$ random density matrix we
write $\rho =U d U^{\dagger}$ and use the
product measure $\mu=\Delta _{\lambda}\times \nu _{H}$. The vector
of eigenvalues $d$, taken according to the Dirichlet measure
(\ref{Dirichlet}), can be obtained from unitary random matrices as shown
in Appendix A. Unitary random rotation matrix $U$ distributed according
to the Haar measure $\nu_H$ is generated with the algorithm presented in
\cite{PZK98}. Random state $\rho$, generated according to a given
product measure,  may be decomposed into a mixture of $N$ pure states
determined by its eigenvectors
\begin{equation}
\rho =\sum_{i=1}^N |\Psi_i\rangle \langle \Psi_i|.
\label{rhoa1}
\end{equation}
Note that the pure states $|\Psi_i\rangle$ are not normalized to
unity, but its norms  are given by the eigenvalues $d_i$.
Expansion coefficients of each of these states
are given by the elements of the random rotation matrix;
$|\Psi_i\rangle= \sqrt{d_i} \{ U_{1i},U_{2i},\dots,U_{Ni}\}$.

There exist many other possible decompositions of the state $\rho$ into
a mixture of $M$ pure states, with $M\ge N$. Let ${\tilde{V}}$ be a
random unitary matrix of size $M$ distributed according to the Haar
measure on $U(M)$. Let
$V$ denotes a rectangular matrix constructed of the $N$ first columns of
${\tilde V}$.  Any such $M \times N$ matrix allows one to write a
legitimate decomposition $\rho'$ of the same state $\rho$
\begin{equation}
\rho' =\sum_{i=1}^M |\phi_i\rangle \langle \phi_i|,
\label{rhoa2}
\end{equation}
  where
\begin{equation}
 |\phi_i\rangle =\sum_{m=1}^N V_{im}|\Psi_m\rangle; ~ ~ i=1,\dots M.
\label{rhoa3}
\end{equation}
Unitarity of the rotation matrix $\tilde {V}$ assures the correct
normalization Tr$\rho'=\sum_{i=1}^M \langle\phi_i|\phi_i\rangle=1$.

 Assume that the composite $N$ dimensional quantum
system consists of two subsystems of the size $N_A$ and $N_B$, such that
$N=N_AN_B$.  It is then convenient to
 represent any vector $|\phi_i\rangle$ (of a non zero
norm  $p_i=\langle\phi_i|\phi_i\rangle$),
by a complex $N_A \times N_B$ matrix $A^{(i)}$, which contains all
$N$ elements of this vector. To describe the reduction of the state
$|\phi_i\rangle $ into the second subsystem we define a $N_B \times
N_B$ Hermitian matrix
\begin{equation}
  B^{(i)} : = [A^{(i)}]^{\dagger} A^{(i)}.
\label{BK}
\end{equation}
Diagonalizing it numerically we find its eigenvalues $\tilde{b}_l^{(i)},
l=1,N_B$. Rescaling them by the norm of the state $p_i$ we get
$b_l^{(i)}=  \tilde{b}_l^{(i)}/p_i$ satisfying
$\sum_{l=1}^{N_B} b_l^{(i)}=1$.
 We compute the entropy of this partition
\begin{equation}
  E_B(|\phi_i\rangle) : = - \sum_{l=1}^{N_B} b_l^{(i)} \ln b_l^{(i)},
\label{entr}
\end{equation}
giving the von Neuman entropy of the reduced state. Entanglement of the
state $\rho'$ with respect to the rotated decomposition (\ref{rhoa2}) is
equal to the average entropy of the pure states involved
\begin{equation}
  E(\rho')  =  \sum_{i=1}^{M} p_i E_B(|\phi_i\rangle),
\label{entr2}
\end{equation}
where $\sum_{i=1}^M p_i=1$.

The entanglement of formation $E$ of the state $\rho$ is then defined as
a minimal value $E_B(\rho')$, where the minimum is taken  over
the set of the decompositions $\rho'$ given by (\ref{rhoa2}),
(compare with the definition
(\ref{formation})). Rotation matrix $V_o$ for which the minimum is
achieved is called the {\sl optimal}. Our task is to find
the optimal matrix in the space of $M$ dimensional unitary
matrices where $M=N,N+1,\dots, N^2$.

We found it interesting to consider also the generalized
entanglement
\begin{equation}
  E_q(\rho')  =  \sum_{i=1}^{M} p_i E_q(|\phi_i\rangle),
\end{equation}
where
\begin{equation}
  E_q(|\phi_i\rangle) : = {1 \over 1-q} \ln \Bigl( \sum_{l=1}^{N_B}
[b_l^{(i)}]^q \Bigr) .
\label{entr3}
\end{equation}
The standard quantity $E_B(\rho)$ is obtained in
 the limit lim$_{q\to 1}E_q(\rho)$.

\subsection{Search for the optimal rotation matrix}

The search for the optimal rotation matrix $V$ has to be performed in
the $M^2$ dimensional space of unitary matrices. Starting with
$M=N$ one has to consider the $16$ dimensional space in the simplest
case of $N=4$. To obtain accurate minimization results in such a large
space one should try to perform
some more sophisticated minimization schemes, for example the stimulated
annealing. Fortunately, the optimal rotation matrix $V_o$
is determined up to a diagonal unitary matrix containing $M$
arbitrary phases.
Therefore, one can hope to get reasonable results with a simple
random walk, moving only then if the entanglement decreases. Performing
only the 'down' movements in the $M^2$ dimensional space one has a
good
chance to land close to the $M$ dimensional manifold defined by
optimal matrices equivalent to $V_o$. This
corresponds to fixing the temperature to zero in the annealing scheme,
and simplifies the search algorithm.

To perform small movements in the space of unitary matrices we will use
$M \times M$ Hermitian random matrices $H$, pertaining to the
Gaussian
Unitary Ensemble (GUE). They can be constructed by independent Gaussian
variables with zero mean and the variance
$(\sigma^{\rm{Re}}_{mn})^2=(1+\delta_{mn})/M$ for the real part and
$(\sigma^{\rm{Im}}_{mn})^2=(1-\delta_{mn})/M$ for the imaginary part of
each complex element $H_{mn}=H_{nm}^*$. We generate random matrix
$H$ and take $W =e^{i\chi H}$
as a unitary matrix, which might be arbitrarily close to the identity
matrix.
Our strategy consists in fixing the initial angle $\chi_0$, performing
random movements of this size and then gradually decreasing the angle
$\chi$.

The detailed algorithm of estimating the entanglement of
formation of a given $N \times N$ state $\rho$ is listed below.

(1) Fix the number $M$ of the components of the decomposition
    (\ref{rhoa2}). Start with $M=N$.

(2) Generate random unitary rotation matrix $V$ of size $M$,
   which defines the decomposition $\rho'$ in (\ref{rhoa2}).
   Compute the entanglement $E=E_B(\rho')$ according to
    (\ref{entr},\ref{entr2}).

(3) Set the initial angle $\chi=\chi_0$.

(4) Generate a random $M \times M$ GUE matrix $H$ and compute
$V'=V\exp(i\chi H)$. Calculate the entanglement $E'$ for the
decomposition $\rho'$ generated by $V'$.

(5) If $E'<E$ accept the move (substitute $V:= V'$ and $E:=E'$)
and continue with the step (4). In the other case repeat the steps
(4-5) $I_{change}$ times.

(6) Decrease the angle $\chi:= \alpha\chi$, where $\alpha <1$.

(7) Repeat the steps (4-6) until $\chi < \chi_{end}$.
   Memorize the final value od the entanglement $E$.

(8) Repeat $L_{mat}$ times the steps  (2-7) starting from a different
initial random matrix $V$.

(9) Memorize the value $E_M$, defined as the  smallest of $L_{mat}$
   repetitions of the above procedure.

(10) Set $M=:M+1$ and repeat the steps (2-9) until $M=M_{max}$.

(11) Find the smallest value of $E_M,
M=N,\dots,M_{max}$. This value $E_{min}=E_{M_*}$ gives the upper bound
for the entanglement of formation of the mixed state $\rho$, while the
size $M_*$ of the optimal rotation $V_*$ may  be considered as
the cardinality of $\rho$.

\subsection{Remarks on estimating the entanglement of formation}

The accuracy of the above algorithm may be easily tested for the case
$N=4$, for which the analytical formula (\ref{conc2}) exist.
Results mentioned in the section IV~B, giving the mean error of the
estimation of the entanglement smaller than
$10^{-7}$, were obtained with the following algorithm parameters:
the initial angle $\chi_0=0.3$, the final angle $\chi_{end}=0.0001$,
the angle reduction coefficient $\alpha=2/3$, the number of iterations
with the angle fixed $I_{change}=25$, the number of realizations
$L_{mat}=3$.
Using relatively slow routines interpreted by {\sl Matlab}
on a standard laptop computer we needed a couple of minutes to get the
entanglement of any mixed state $\rho$. Although we performed test
searches with
$M=4,5,\dots,8$ the optimal rotation was always found for $M=N=4$.

The same algorithm was used for random states with $N=2 \times 4=8$. In
this case the simplest search with $M=N=8$, performed in $64$
dimensional space, requires much more computing time.
It depends on all parameters characterizing the algorithm; one may
impose therefore an additional bound on the total number $I$ of
generated random matrices $V'$. To estimate the volume of the bound
entangled states we performed the
above algorithm only for the states with positive partial transpose.
Setting the final angle at $\chi_{end}=0.0002$
we obtained in a histogram $P(E)$ a flat local minimum at $E_{m} \sim
0.0003$. The minimum is located just right to the singular peak at even
smaller values of
$E$, corresponding to separable states.
The cumulative distribution $P_c=\int_{E_m}^{\infty} P(E)dE$ was found
not to be very
 sensitive on the position of the minimum  $E_m$.  We could, therefore,
set the cut-off value $E_c$ to the position of minimum $E_m$, and
interpret the quantity $P_c$ as the relative volume of the bound
entangled states.  In the computations described in section IV~C
 we took  $\chi_0=0.3$,
$\alpha=2/3$, $I_{change}=25$ and  $L_{mat}=5$,
and for $M=N=8$ obtained the mean number of iterations
$\langle I \rangle$ of the order $5\times 10^4$.

The other possibility
to distinguish the bound entangled states from the separable
states consists in studying the dependence of the obtained upper bound
on the entanglement $E$ on the total number $I$ of iterations performed.
Numerical results obtained for the separable states show that
 $E$ decreases with the computation time not slower than $E(I)= a/I$.
Assuming a similar effectiveness of the algorithm for the nonseparable
states (with a non zero  entanglement of formation $E_{form}$),
we have $E(I)=E_{form}+b/I$. This allows us to design a simple
auxiliary criterion: the state $\rho$ is separable if
(for all realizations of the random walk starting from the
different matrices $V$) for sufficiently large number of iterations $I$
one has $E(I)< E(I/2)/2$. If this condition is not
fulfilled the state $\rho$ can be regarded as entangled.
Using this method we obtained the estimation for the volume of bound
entangled states similar to $P_c$.

\end{document}